\documentclass[12pt,preprint]{aastex}

\shorttitle{Direct Impact Accretion AM CVn Binaries}
\shortauthors{Dolence, Wood, and Silver}

\newcommand{\FITDisk}{{\sc FITDisk}}

\begin{document}

\title{SPH Simulations of Direct Impact Accretion in the Ultracompact AM CVn Binaries}

\author{Joshua Dolence\footnote{Current address: Dept. of Astronomy, University of Illinois, Urbana, IL 61801}, Matt A. Wood, and Isaac Silver}
\affil{Department of Physics and Space Sciences and SARA Observatory, Florida Institute of Technology, Melbourne, FL 32901}
\email{dolence2@astro.uiuc.edu, wood@fit.edu, isilver@fit.edu}

\begin{abstract}
The ultracompact binary systems \object[RX J1914.4+2456]{V407 Vul} (RX J1914.4+2456) and \object[RX J0806.3+1527]{HM Cnc} (RX J0806.3+1527) -- a two-member subclass of the AM CVn stars -- continue to pique interest because they defy unambiguous classification.  Three proposed models remain viable at this time, but none of the three is significantly more compelling than the remaining two, and all three can satisfy the observational constraints if parameters in the models are tuned.  One of the three proposed models is the direct impact model of \citet{MS2002}, in which the accretion stream impacts the surface of a rapidly-rotating primary white dwarf directly but at a near-glancing angle.  One requirement of this model is that the accretion stream have a high enough density to advect its specific kinetic energy below the photosphere for progressively more-thermalized emission downstream, a constraint that requires an accretion spot size $\sim$$1.2\times10^5\rm\ km^2$ or smaller. 
Having at hand a smoothed particle hydrodynamics code optimized for cataclysmic variable accretion disk simulations, it was relatively straightforward for us to adapt it to calculate the footprint of the accretion stream at the nominal radius of the primary white dwarf, and thus to test this constraint of the direct impact model.  We find that the mass flux at the impact spot can be approximated by a bivariate Gaussian with standard deviation $\sigma_{\phi} = 164\rm\ km$ in the orbital plane and $\sigma_{\theta} = 23\rm\ km$ in the perpendicular direction. The area of the the $2\sigma$ ellipse into which $\sim$86\% of the mass flux occurs is roughly 47,400 $\rm km^2$, or roughly half the size estimated by \citet{MS2002}.  We discuss the necessary parameters of a simple model of the luminosity distribution in the post-impact emission region.
\end{abstract}


\section{Introduction}

Understanding the exact physical nature of the ultracompact binaries (UCBs) \object[RX J1914.4+2456]{V407 Vul} (RX J1914.4+2456) and \object[RX J0806.3+1527]{HM Cnc} (RX J0806.3+1527) has been the goal of extensive studies since these two closely related systems were discovered.  
The observational results tightly constrain the universe of possible solutions \citep{Cro2004,Bar2007}:  

\begin{enumerate}
\item The ``on/off'' X-ray light curves folded on phase look nearly identical for the two systems, rising steeply to maximum, declining by half over the width of the pulse, and then dropping steeply again to zero 
\item The single observed and so presumed orbital periods are 321.5 s (5.36 min) for HM Cnc \citep{Mot1995,Cro1998,Ram2002} and 569.4 s (9.49 min) for V407 Vul \citep{Ram2000,Ste2006}, 
\item The observed periods are decreasing with time \citep{Str2002,hak03,Str2004,Ram2005,Ram2006}, consistent with the rate predicted for angular momentum losses by gravitational radiation,
\item There is an apparent phase difference in the optical and X-ray light curves in both systems in which the optical light curves lead the X-ray light curves by $\sim$0.2 in phase \citep{Bar2007},
\item No hard X-rays are detected, 
\item There is long-term variability in the mean luminosities of the systems \citep{isr99,Ram2002,isr02},
\item Flickering, if present in the optical light curves, is of small amplitude, and suggests that the optical light is not dominated by radiation coming directly from the accreting region \citep{Bar2007}.
\end{enumerate}

Of the several models proposed to explain the early observations of these systems, three remain viable at this time: (i) the polar model, (ii) the direct impact (DI) model, and (iii) the unipolar inductor (UI) model.   In all three of the currently viable models both the primary and secondary star are thought to be white dwarf stars, but only for the polar and DI models does the secondary fill it's Roche lobe.  Each of the three can explain the observations with varying degrees of success, but some fine tuning is required,  and in the case of UI the physical process is unfamiliar and unproven in a stellar context. 

Historically, the system V407 Vul was discovered first and suggested to be a double degenerate ``soft'' intermediate polar (IP) with a white dwarf spin period of 568 s \citep{Mot1995}.  \citet{Cro1998} noted that no other known IP displayed {\it zero} X-ray luminosity for $\sim$1/2 the observed period, and they proposed that the system was instead a double degenerate analog of the polars.  
HM Cnc was identified in the following year \citep{isr99,beu99}, and in 2002 two separate groups suggested independently that this system is a near twin of V407 Vul \citep{Ram2002,isr02}.  In contrast, the binary ES Cet with an orbital period of 10.3 min, only slightly longer than 9.5 min for V407 Vul, shows clear signs of an accretion disk, and is a typical AM CVn binary \citep{war02,Cro2004}.  The IP model is unlikely to apply to these systems, in that the pulse shape of the X-ray light curve, the lack of hard X-rays and emission lines, and the infrared colors that seemingly exlude a main-sequence secondary are all difficult to explain within the IP model \citep{Cro2004}.  We do not discuss this model further, but the interested reader may wish to consult \citet{Nor2004}.

The polar model for the V407 Vul stars proposes that a Roche-lobe filling secondary transfers He-rich plasma to the magnetized and synchronized primary star, channeled by the magnetic field onto the nearest magnetic pole which is fixed in the co-rotating frame.  Orbital motions take the accretion spot and column in and out of view, producing the X-ray lightcurves.  The phase-offset optical variations are thought to arise from reprocessing on the face of the secondary star. Problems with this model include that emission lines and polarization are common in polars but not observed here, and mass transfer from a lower-mass secondary to a higher-mass primary should increase the orbital period.  In regards to this last point, however, two separate groups \citep{deloye06,dantona06} have shown that secondaries that are still contracting and non-degenerate can yield decreasing orbital periods while the mass transfer is in early phases, and can exist in this state for $10^3$ to $10^6$ years.  This is sufficient time for the requirements of observed space densities based on population synthesis models \citep{nelemans01}. 

The DI model proposes that the two stars are close enough that the accretion stream impacts the surface of the primary white dwarf directly \citep{MS2002} because the stellar radius is larger than the ``periastron'' of a ballistic stream originating at the inner Lagrange point L1.  \citet{MS2002} argue that in the DI model the accretion stream will be sufficiently narrow (width $\sim10^{-4} R_\odot$) that the density of accreted material will be high enough to penetrate below the photosphere and thermalize to soft X-rays, similar to the blob accretion model proposed for some polars \citep{kui82,fkl88,king00}.  For this to apply in the DI model, the surface area of the accretion spot must be a fraction $f\sim10^{-4}$ of the surface area of the star or smaller, which translates to an area smaller than $\sim1.2\times10^5\rm\ km^2$.  The initial motivation for this project came from Pasi Hakala, who suggested to one of us (MW) that it would be useful to use our SPH code to determine accurately the accretion spot size.  If a hydrodynamic simulation were to find that the footprint of the accretion stream is substantially larger than the \citet{MS2002} estimate, then that would be a serious problem for the DI model.  Of course, it is a one-sided test, and passing the test does not mean that the DI model is correct, just consistent with the physical constraints of the blob accretion model.

The unipolar-inductor (UI) model proposed by \citet{Wu2002} and discussed most recently in the context of V407 Vul and HM Cnc by \citet{dal06,dal07} and \citet{ram07} is the stellar analog of 
behavior observed in the Jupiter-Io system.  As Io moves through Jupiter's magnetic field, charge separation occurs, and because the environment is a weak plasma, an emf is established from the far point of Io as observed from Jupiter, along the magnetic field lines in both directions to Jupiter's atmosphere in both hemispheres, and from the planet back to the near side of Io \citep{glb69}.  Resistive heating in the region between the two spots in each pair resulting from the current flow makes these regions luminous, as was observed by the {\it Hubble Space Telescope} \citep{clark}.  \citet{Wu2002} suggested the model scaled to HM Cnc and V407 Vul could explain the observations -- certainly a fascinating and clever idea, even if eventually determined not to be applicable in these stars.  

In a stellar context, the rate of energy dissipation in the UI model can be of order $L_\odot$ or higher, and with the predicted hot spot area $\sim8\times10^{4}\rm\ km^2$, the \citet{Wu2002} model predicts a 50 eV temperature consistent with the 55 eV observed from ROSAT and ASCA spectral fits \citep{Ram2000}.  The hotspots are on the plane that passes through both the line of centers and the orbital rotation axis, and the source of the optical light modulation is again assumed to be reprocessed radiation from the heated face of the secondary.  An important point in this model is that the secondary is not Roche-lobe filling, and there is no significant mass transfer. To first order the timescale for orbital evolution is that which gravitational radiation predicts, consistent with the observed rates of period change \citep{Str2002,hak03,Str2004,Ram2005,Ram2006}.  \citet{ram07} note that the UI model predicts radio emission, which they detected at the 5.8-$\sigma$ level along the line of sight to HM Cnc, but which was not detected in observations of V407 Vul.  The major problem with this model is the phasing of the optical versus X-ray light curves, because the X-ray spots are on the meridian facing the secondary, this model predicts a phase offset of $\sim$$0.5$, whereas the observed phase offset is $\sim$$0.2$ for both V407 Vul and HM Cnc \citep{Bar2007}.  In addition, \citet{MN2005} suggest there are serious problems with synchronization timescales, but \citet{dal07} argue that Marsh \& Nelemans result rests on an estimate of the luminsoty of V407 Vul which is too high by a factor of $\sim10^2$.  However, for the UI model to explain both systems, \citet{dal07} find that the primary white dwarf in HM Cnc rotates nearly synchronously but slightly slower than the orbital period, whereas the primary in V407 Vul must have a rotation frequency which is  roughly 10\% higher than the orbital frequency.  The authors do not suggest how prior binary evolution without mass transfer can yield such a rapidly-rotating primary.

In all three of the models above, the on/off behavior of the X-ray light curve has been interpreted to mean that the X-rays must be emitted from a geometrically compact region \citep{Cro1998,Cro2004}.  For the UI model, this region is the footprint of the magnetic field lines which carry current from the secondary (non-magnetic) to primary (magnetic) star.  \citet{Bar2005} analyzed this model and were able to place strict constraints on the allowable parameter space based on geometrical arguments involving the size of this footprint.  Here, we present the results of a hydrodynamic simulation of V407 Vul to estimate the area of the footprint of the accretion stream, to determine if the calculated spot size is small enough to satisfy the DI model requirement.  In \S 2 we discuss the changes made to our SPH code \FITDisk\ to pursue this project and the system parameters assumed.  In \S 3 we discuss the characteristics of the simulated accretion stream, and in \S 4 we present our primary results. In \S 4 we discuss our results in the context of the existing literature on V407 Vul and HM Cnc, and in \S 6 we present our conclusions.

\section{Numerics}
\subsection{The Code}

We use the method of smoothed particle hydrodynamics, a Lagrangian-based technique that is well-suited for astrophysical problems with unusual geometries and vacuum outer boundary conditions \citep{lucy77,monaghan92,mon05}.  Our code \FITDisk\ was devolped to model accretion disks in cataclysmic variables \citep{simpson95,SW1998,wea05,wds06,wb07}, and the details of the numerics can be found in those publications.  In brief, \FITDisk\ solves in the inertial frame the equations of compressible, viscous hydrodynamics subject to the external gravitational field of two point-mass stars.  Particles all have the same effective radius (smoothing length $h$) but can have different time step sizes as dictated by local conditions.  In our typical simulation runs, we build up a disk by injecting $\sim$$2000$ particles per orbit at the L1 point until we reach the target number of particles in a particular run ($\sim$$10^5$).
A particle is considered {\it accreted} onto $M_1$ when it is closer than the specified radius of the star.  When this occurs (or accretion onto $M_2$ or ejection from the system), a replacement particle is immediately injected at the L1 point.  Thus, the simulation in time settles into a dynamical equilibrium state.

Our goal with the current project is to test the requirement of the DI model of \citet{MS2002} that the accretion spot on the surface of $M_1$ has an area $A \lesssim1.2\times10^5\rm\ km^2$ such that the stream density is sufficient to advect and release the stream kinetic energy below the photosphere, which is required to avoid hard X-ray emission.  For this preliminary test, then, our main requirement is that the accretion stream behave as a fluid from the L1 region to the surface of the primary.  We keep track of where the trajectories of the simulation particles intersect the mathematical surface of the primary, and use the resulting integrated mass-flux distribution to estimate the effective area of the acccretion spot.  Our goal here is not to model in detail the hydrodynamics of the impact region -- we are only beginning to address that far more ambitious calculation, and it will be the topic of a future publication.  Our current results can be considered to provide the minimum estimated effective area for the accretion spot, as all physical effects we've neglected could only act to increase the effective area. 

To explore this problem, we made two substantial changes to our base code.  First, we converted the code to make the calculations in the co-rotating frame instead of the inertial frame used in \FITDisk\ up to this point, and second we implemented a new particle injection scheme which gives a far more physical model of the flow dynamics through the inner Lagrange point L1.

Our original code calculates in the inertial frame, which generally simplifies the analysis and visualization steps, but for this project the analysis is simplified if the calculations are completed in the co-rotating frame with the implementation of coriolis forces.  We establish a Cartesian system, with the center of mass at the origin.  The $x$-axis is along the line of centers of the two stars, with the donor star orbiting in the $+\hat{y}$ direction and the orbital angular momentum vector in the $+\hat{z}$ direction.   Within this system, the stars, stream, and impact spot are stationary features.  Therefore, the approximate point of impact for a given particle can be easily calculated by linear interpolation along its trajectory between its positions at time steps $t^{n-1}$ and $t^{n}$ bracketing the primary surface radius.   Also, in a steady state it becomes meaningful to examine the cumulative distribution of the particle mass flux as a function of position on the surface of the primary as a means to estimate the effective accretion spot size.

The mass transfer stream is created {\it ab initio} by a Monte-Carlo-based particle injection technique.  Particles are injected at random locations in a small region near the L1 point.  In the original code, the region is a box located on the primary side of the L1 point, and the L1 point is at the center of one face.  Particles are injected with velocity vectors calculated assuming sound-speed flow through L1, and we naturally recover the trajectories of \citet{LS1975}.  In order to resolve the accretion spot, we used an effective scaled particle size of 10 km, or $h\sim10^{-4}$ in simulation units where the stellar separation $a\equiv1$.  With particles this small, we found the particle trajectories were ballistic and not behaving as a fluid (cf. Cash 2002, 2004).  The solution we arrived at was to move the injection region to the {\it secondary} side of L1, and to implement a reflecting boundary plane approximately $30h$ from the L1 point. We inject particles with zero velocity in this region, and hence create a crude pressure-supported atmosphere that naturally leads to a physical flow distribution through the L1 region into the Roche potential of the primary.  To avoid spuriously generated sound waves when injecting particles, we find that preferentially injecting particles near the base of the atmosphere, where the density is significantly higher, is an effective technique since viscous interactions tend to quickly damp any excited modes.  This technique provides the most accurate flow dynamics through the L1 region obtainable within the SPH approach.  

The Reynolds number is defined as
\begin{equation}
Re = {v_sL\over \nu},
\end{equation}
where $v_s$ is the mean fluid velocity $L$ is the characteristic size of the region, and $\nu$ is the dynamic fluid viscosity. We can estimate the Reynolds number in the L1 region in our simulation by using the results of \citep{mon97} who finds that the kinematic shear viscosity within the SPH formalism can be approximated as
\begin{equation}
\nu = {15\over112}Kv_{\rm sig}h
\end{equation}
where $K$ is a constant ($K\sim0.5$) and $v_{\rm sig}=c_i + c_j$ is the signal velocity between particles $i$ and $j$.  As noted above, we place our reflecting surface $30h$ behind the mathematical L1 point, which is roughly the width of the stream at the L1 point (see Figure 1), and we use this as the characteristic size $L$ (in pipe flow, the width of the pipe is taken as $L$).  The mean fluid velocity is $v_s\sim0$ at the reflecting wall, and $v_s\sim c_s$ as the flow passes through the L1 point, so we use $v_s=c_s/2$ in the estimate of the Reynolds number.  Then from equations (1) and (2) and using
$K\sim 1/2$, $v_{\rm sig}=c_i+c_i\sim 2c_s$, and $L\sim 30h$, we have
\begin{equation}
Re = {v_sL\over \nu} \sim {(c_s/2)(30h) \over {15\over 112}(1/2)(2c_s)h} \sim 112.
\end{equation}
This value of the Reynolds number is roughly 3-4 orders of magnitude lower than the typical transition to turbulent flow, so we expect the fluid near the L1 region to be non-turbulent.  Furthermore, we note that \citet{del07} in their recent study of the structure and evolution of the mass donating secondary stars found that the atmospheric structure of these stars just after mass transfer is initiated is characterized by radiative stability at the photosphere, a thin and weak sub-photospheric convection zone, and then another radiative zone that extends to the core.

As discussed above, particles that are accreted onto the primary are immediately reinjected in the L1 region so that a constant number of particles is maintained in the simulation, and a dynamical equilibrium flow is established.  For the simulation presented here, a steady-state mass transfer rate is established less than one orbit after the number of particles reaches its preset maximum value.

\subsection{The Model}

For the current study, we adopt the system parameters suggested by \citet{MS2002} for V407 Vul: $P_{\rm orb}=9.5$ min, $M_{1}=0.5 M_{\sun}$, $M_{2}=0.1 M_{\sun}$, and $R_{1}=0.0144 R_{\sun}$.  The orbital separation is calculated from Kepler's third law and is taken to be constant.  Distances are scaled so that the orbital separation is unity in dimensionless system units.  Because the mass of the SPH particles formally drops out of the equations (see SW), no particular mass transfer rate is assumed a priori.  The total number of particles is set to 86,400. Particles are given an initial temperature of 20,000~K, appropriate for the heated face of the donor star \citep{MS2002}.  The viscosity coefficients for the \citet{lattanzio86} artificial viscosity prescription are $\alpha=1.0$ and $\beta=0.5$, although viscosity is expected to play a negligible role in the dynamics of a stream of plasma in free-fall.  An ideal equation of state is used, with $\gamma=1.01$.  The smoothing length was set to a value expected to be small with respect to the impact spot, which in our case was $h=1.1493 \times 10^{-4}$ in simulation units, or $h \approx 10\rm\ km$.  Tests were done with slightly larger $h$ and the result did not change significantly.  With a smoothing length of this size, we were forced to use 6,000 time steps per orbit with three levels of refinement (i.e., $\delta t_{0}=P_{rm orb} / 6000$, $\delta t_1 = \delta t_0/2$, or $\delta t_2=\delta t_0/4$) as described more fully in SW.  We note that the time-of-flight $t_f$ from L1 to impact can be estimated using Kepler's third law with $a'\sim a/2$, and then the time of flight from L1 to impact is roughly $t_f \sim (1/2)(a'/a)^{3/2} P_{\rm orb} \sim P_{\rm orb}/5 \sim 1$-2 min, where the factor of $1/2$ is the result of the particle completing only half the orbit.
 
While we make an attempt to accurately calculate the stream profile from the L1 region to impact spot, we make several simplifying assumptions.  First, we assume that rotation of the secondary and surface flow on the secondary star were negligible.  This is likely to be a reasonable assumption as the secondary is almost certainly co-rotating at the binary period.  We further assume for the purpose of these calculations that in the absence of magnetic fields the accretion spot size will not be greatly enhanced by interactions with the fluid already at the impact point, and we assume that each particle impacts at the point where it's trajectory takes it through the radius of the primary.  This assumption should hold to a good approximation, because the fluid in the accretion stream is highly supersonic when it impacts the surface, and in the absence of significant surface magnetic fields, the surface fluid is unable to apply any force on the incoming fluid before impact with the exception of (negligible) radiation pressure.  Finally, in all calculations, radiative effects and magnetic forces are explicitly ignored.

\subsection{Analysis}

As described above, our technique lends itself to examining the cumulative distribution of particles over a long span of simulation time.  Once our simulations have reached dynamical equilibrium, the probability distribution function of particles is constant with time.  By utilizing a Monte-Carlo approach to particle injection, we ensure that the particular positions of the particles changes over the course of the simulation.  We further ensure a random selection of particles by sampling the simulation every $60^{th}$ time step so that a particular particle does not contribute to the number density of particles in any location more than once.  Note that particles within the Roche-lobe of the secondary could potentially be sampled many times by this technique, but this is of no concern here.  With these aspects in mind, a two-dimensional Eulerian mesh (with small thickness) is created that slices through a region of interest and the number of particles that lie within each grid zone is summed over all of the time steps included by the above criterion.  In this way, we effectively increase the resolution of our results and reduce the noise inherent to the SPH technique (see also Wood et al.\ 2005).

To characterize the impact spot, many of the above ideas apply.  The key difference is that we record the impact position of every particle accreted, as opposed to sampling every $60^{th}$ time step.  This is because each particle can only cross the surface of the primary one time by design, and so the Monte-Carlo procedure for particle injection, coupled with the significant mixing within the Roche-lobe of the secondary, serve to randomize the particle impact points.  Also, since the impact spot lies on the curved surface of the primary, we lay our mesh on this curved surface and utilize the angular position of the impacts to bin them into grid zones.  While this creates zones of different linear dimensions, the effect is negligible since the impact spot lies in a very small region centered on the orbital plane.  Once the particles have been binned, the relative positions of the grid zones are converted back to linear dimensions so that the distances presented below are actually arc lengths.  With this complete, the distribution is fit with a bivariate Gaussian using a nonlinear least squares analysis.

\section{Stream Characteristics}

In this section, we briefly recount the results of several previous studies which explored the dynamics of the mass transfer stream in accreting binary stars.  We focus on two regions, namely the L1 region and the impact spot on the primary, the former primarily for validation of our technique.

\subsection{The L1 Region}

In the L1 region, \citet{LS1975,LS1976} showed via semianalytic methods that the density cross section of stream should Gaussian, and decrease in overall density as $r^{-1}$ where $r$ is the distance from the L1 point along the center of the stream.  They also showed that the stream is deflected by an angle $\theta \approx 20.8 \degr$ from the line connecting the centers of the stars (for mass ratio $q=0.2$).  Later numerical work by  \citep{numsurf}, which accounted for the surface flow on the secondary, found that the stream was approximately Gaussian about its center (slightly bow shaped convex towards the trailing side), decreased as $r^{-0.8}$, and had a deflection angle smaller by roughly a factor of two compared to Lubow \& Shu.

\subsection{The Accretion Stream Impact Region}

As the stream material accelerates toward the the primary, it is stretched by tidal forces which act to narrow the stream, especially in the $z$-direction.  This was predicted by \citet{LS1976}, who showed that the vertical stream width at closest approach to the primary should be of order the hydrostatic scale height of a disk at that same radius.  The pressure scale height of a hydrostatic disk at $r=R_{1}$ from the center of the primary is approximately given by
\begin{displaymath}
H_{p} \sim \left(\frac{kTR_{1}^{3}}{GM_{1}m_{p}} \right) \sim 6 \times 10^{-4} 
\end{displaymath}
in simulation units.  \citet{LS1975} find that the stream width in the orbital plane should be roughly a factor of five larger than the vertical width at closest approach, or $\sim 3 \times 10^{-3}$ in simulation units.  While in our case the stream is intercepted by the surface of the primary before it reaches its closest approach, we take these values as approximate, and indeed find our results to be in good agreement with those of \citet{LS1975,LS1976}.

\section{Results}

In Figure~1, we show the trajectory of the stream by plotting the stars and the SPH particles at a single time step viewed at orientations perpendicular and parallel to the orbital plane.  As previously noted, the impact spot on the primary white dwarf is on the far side as viewed from the secondary \citep{MS2002}.

As a proof of concept, we begin by analyzing our results for the flow in the L1 region.  Our results are shown in Figure 2 as density contours of a slice centered on the orbital plane.  We find that the stream is indeed approximately Gaussian distributed about its center line, decreases in density as approximately $r^{-0.5}$, and is deflected by an angle $\theta \approx 21 \degr$, in reasonable agreement with the \citet{LS1975,LS1976} results. 

In Figure~3, we show contours of constant mass flux through the surface of the primary and the best fit obtained via nonlinear least squares analysis using a bivariate Gaussian function.  For the purposes of fitting, a coordinate system with origin at the center of mass flux is employed.  The mass flux, $\Phi$, as found by our model is of the form
\begin{equation}
\Phi(s_{\phi},s_{\theta}) = N_\Phi \exp \left[-\frac{1}{2} \left(\frac{s_{\phi}^2}{\sigma_{\phi}^{2}} + \frac{s_{\theta}^2}{\sigma_{\theta}^{2}} \right) \right]
\end{equation}
where $N_\Phi$ is a constant scaling factor, $s_{\phi}=\phi R_{1}$ is the arc length in the azimuthal direction, $s_{\theta}=\theta R_{1}$ is the arc length in the polar direction, and $\sigma_{\phi}$ and $\sigma_{\theta}$ are the standard deviations in these directions.  In these coordinates, $s_{\phi}$ increases in the $-\hat{x}$ direction and $s_{\theta}$ increases in the $+\hat{z}$ direction.  The result shown in Figure 3 is normalized to unity at the origin.  The best fit model gives $\sigma_{\phi}=0.0019$ and $\sigma_{\theta}=0.00026$ in simulation units, or $\sigma_{\phi} = 164\rm\ km$ and $\sigma_{\theta} = 23\rm\ km$ in physical units.  The eccentricity of the resulting ellipse is $e = \sqrt{1-(\sigma_\theta/\sigma_\phi)^2}\approx 0.99$.

\section{Discussion}

Looking again at Figure~1, it is apparent that the supersonic accretion stream impacts the white dwarf surface with a nearly horizontal trajectory.  The altitude angle of the stream as viewed from the impact point is $\lesssim$$20^\circ$).  A very approximate mental picture of the situation can be achieved by considering a narrow, high-velocity jet of water impacting the surface of a pool with a similarly shallow altitude angle, and in fact the experiment is so easily available we recommend the interested reader actually try it.  In the stellar case, the shock-heated surface fluid will be promptly carried away from the impact point, and the accretion stream will always be impacting fresh material that is entrained from the surrounding $T\sim T_{\rm eff}$ atmosphere.  There will be strong radial flows towards the core of the impact spot which will serve to advect on a somewhat slower time scale the specific kinetic energy of the accreting material in the lower-density outer volume of the stream. 

We characterize the shape of the impact region as an ``$N$-$\sigma$'' ellipse with semi-major and semi-minor axes given by for example $a=\sigma_\phi$ and $b=\sigma_\theta$ for the 1-$\sigma$ case, and where $N$ need not be an integer.  If we now assume our coordinate system is centered on the centroid of the mass-flux distribution, and that the distributions along both the $\phi$ and $\theta$ axes are normal gaussians, then 
the constraining equation for the bivariate normal density distribution for the $N$-$\sigma$ contour is (for example, see Kawka et al. 2004, \S4.2)
\begin{equation}
\left({s_\phi\over N\sigma_\phi}\right)^2 + \left({s_\theta\over N\sigma_\theta}\right)^2 < 1
\end{equation}
or
\begin{equation}
\left({s_\phi\over \sigma_\phi}\right)^2 + \left({s_\theta\over \sigma_\theta}\right)^2 < N^2.
\end{equation}
The joint probability of finding an object within the $N$-$\sigma$ ellipse is then the integral of the probability density function inside the ellipse:
\begin{equation}
P(N\sigma) = {1\over 2\pi}\displaystyle\int_{-N}^{N} e^{-u^2/2} \displaystyle\int_{-\sqrt{N^2-u^2}}^{\sqrt{N^2-u^2}} e^{-v^2/2}\ dv\,du,
\end{equation}
where $u=s_\phi/\sigma_\phi$ and $v=s_\theta/\sigma_\theta$.
This yields for the 1-, 2-, and 3-$\sigma$ countours:
\begin{eqnarray}
P(1\sigma) &=& 0.36\\
P(2\sigma) &=& 0.86\\
P(3\sigma) &=& 0.99.
\end{eqnarray}
Thus approximately 86\% of the mass in the accretion stream impacts a spot of area $A_{2\sigma}=47,400\rm\ km^2$, or roughly a fraction $f_{2\sigma}=3.8 \times 10^{-5}$.  The 3-$\sigma$ profile (99\% of the mass flux) yields an area $A_{3\sigma}=107,000\rm\ km^2$, or roughly a fraction $f_{3\sigma}=8.5 \times 10^{-5}$, or about the size of Iceland on a white dwarf of radius $\sim$$1.5R_\oplus$.
Our numerical results confirm that the footprint of the accretion stream is remarkably compact, and is consistent with the requirement of the DI model we set out to test.  

We can determine whether the accretion stream penetrates below the photosphere by comparing the ram pressure of the accretion stream to the gas pressure in the white dwarf atmosphere. If we take a conservative estimate for the accretion rate $\dot M=10^{-10}\rm\ M_{\sun}\rm\ yr^{-1}$ \citep{del07}, then we find that the distribution of the stream ram pressure in the radial direction has the same form as the mass flux given above, with a peak value of $3.4 \times 10^{9}\rm\ dynes\rm\ cm^{-2}$, as shown in Figure~4.  White dwarf model atmosphere calculations \citep{vf92,kawkaetal07} for a $T_{\rm eff}\approx20,000$ K, $\log g = 7.75$ white dwarf yield gas pressures at optical depths $\tau = 1$ and $10$ of $P_{\rm gas} \approx 3\times10^5$ and $2 \times 10^{6}\rm\ dynes\rm\ cm^{-2}$, respectively.  Thus, $P_{\rm ram}\gg P_{\rm gas}$, and more than 99\% of the mass flux should advect below the photosphere where the bulk of its energy can be thermalized as in the blob accretion model proposed for some polars \citep{kui82,fkl88,king00}.  

\citet{MS2002} suggested that because the specific kinetic energy of the accretion stream enters at such a shallow angle to the surface, there should exist a high-luminosity region just downstream of the accretion impact point that emits in soft X-rays, and further downstream in the prograde equitorial flow one would expect the spectral energy distribution to soften further towards mean stellar effective temperature.  \citet{Ram2005} report spectral softening as the flux declines towards the off-phase in V407 Vul.  The mean effective temperature of HM Cnc is estimated by \citet{Bar2007} to be in the range 18,500 K to 32,400 K (and probably similar for V407 Vul but currently unmeasurable because of the G star which is conincident with that binary).  We assume $T_{\rm eff}\approx 25,000$ K for the purpose of some simple estimates discussed below.  
The peak optical radiation will thus lead the X-ray peak in orbital phase.  Recent observations \citep{Bar2007} indicate that both systems HM Cnc and V407 Vul are characterised by the optical pulses leading the X-ray pulses by about 0.2 in orbital phase, and that earlier claims \citep{isr03,isr04} that the optical and X-ray pulses were antiphased were the result of incorrect timing corrections for the optical data.  

We can crudely model the surface luminosity distribution using 4 zones: the first three are the accretion spot, the soft X-ray spot, the optical spot -- each of which is assumed to be uniformly illuminated -- and the forth is the remaining stellar photosphere.  All three of the spots will be highly-elongated features, and very roughly we approximate them as ellipses with a common focus (the center of the impact spot) and eccentricity\footnote{We note that this approximation is simply the assumption that the area of the spots are proportional to the square of the length -- the results would be essentially the same for scaled rectangular regions where all 3 aligned on one edge.} (see Figure 5).  As the newly-accreted and shock-heated gas flows along the equator, shear mixing and high levels of turbulence will effectively spread the width of the accretion luminosity flow in latitude as the material propogates in longitude away from the impact spot.  

Because the X-ray flux is observed for 60\% of the orbital period, the soft X-ray emitting region must have a linear extent $\sim$10\% the circumference of the star, or equivalently 0.1 in orbital phase.  This corresponds to a physical length of $s \approx 0.1(2 \pi R_{1})=6100\rm\ km$, which is roughly 9 times longer than the full length of the 2-$\sigma$ accretion spot discussed above.  Measuring from center of the accretion spot, the distance to the center of the X-ray spot will be roughly 0.05 in phase, or 3050 km. Because for high eccentricity ellipses the distance from the focus to the center of the ellipse is $ea\approx a$, for a uniformly-illuminated optical spot to lead the X-ray spot by 0.20 in phase its center would be at a phase angle of 0.25 as measured from the accretion spot, which then implies that the optical spot extends for $\sim$0.5 in phase around the primary white dwarf (Figure 5).  In the limit where the area scales as the square of the ellipse length minus the area of the next smaller ellipse, the ratio of areas of the 3 zones is then 1:80:1920, and the ratio of the areas of the X-ray to optical spots is 1:24.  In the limit where the areas of the zones scale simply as the length of the zones (constant width of order the impact spot width), the ratio of areas for the 3 zones would be approximately 1:9:45.  We note that the real emission regions will have monotonically-decreasing radiation fluxes as a function of distance from the impact spot, which then means that the above estimates are lower limits to the linear extents of the X-ray and optical zones.

\citet{Bar2007} reports that the peak-to-peak amplitude for HM Cnc is roughly 25\% in the optical.  For a 2-$\sigma$ spot viewed in the center of the stellar disk, this implies a spot temperature which is some 16\% higher than that of the surrounding photosphere ($\sim$29,000 K).  At an inclination of $i=45^\circ$, the spot temperature needs to be some 40\% higher than that of the surrounding photosphere ($\sim$35,000 K).  The amplitudes in V407 Vul are diluted by the the G-star which is along the same line of sight and thus not known unambiguously. 

The DI model appears to be able to account for the observations of both systems V407 Vul and HM Cnc, if we accept the results of \citet{deloye06} and \citet{dantona06} that a contracting lower-mass secondary can yield a negative rate of period change and a steady accretion rate for a sufficient length of time that the objects' space density is high enough to be consistent with the observations.    
\citet{Cro2004} suggests that the character of the X-ray light curves of these two systems is difficult to understand in the context of the DI model, where they propose that within the DI model one would expect an X-ray light curve that showed a slow rise to maximum and a steep decline, opposite to that observed.   \citet{Ram2005} find that the spectral softness varies over the orbital phase, and interpret this to mean that the X-ray emitting region has some temperature structure.  More work -- both observational and theoretical -- will be required to understand the nature of the luminosity distribution on the surface of the white dwarf primary within the DI model, and whether there is a unique solution that explains the set of observations.

\section{Conclusions}

We have shown that for system parameters appropriate to V407 Vul, the impact spot size estimated with a hydrodynamics simulation is small enough to be consistent with the requirement of the DI model of \citet{MS2002}.  In summary, we find:

\begin{enumerate}
\item The impact spot calculated for the system parameters of V407 Vul is very compact, with an effective area only 47,400 $\rm km^2$ or a fraction $f=3.8 \times 10^{-5}$ that of the of the primary star assuming the 2-$\sigma$ contour through which 86\% of the mass flows.  The 3-$\sigma$ profile (99\% of the mass flux) encloses an area $A_{3\sigma}=107,000\rm\ km^2$, or roughly a fraction $f_{3\sigma}=8.5 \times 10^{-5}$ of the stellar surface area.
The mass flux at the spot is characterized by a bivariate Gaussian distribution with standard deviations $\sigma_{\phi} = 164\rm\ km$ and $\sigma_{\theta} = 23\rm\ km$.  

\item Within the DI model, the accretion stream impacts the surface nearly horizontally (altitude $\lesssim$20$^\circ$) and supersonically.  The specific kinetic energy is advected and thermalized below the photosphere to be reemitted downstream from the impact point in a very elongated region along the equator and this equitorial flow is characterized by decreasing localized temperature and therefore decreasing spectral energy distribution.

\item The X-ray emitting region must extend for $\sim$10\% of the equitorial circumference to account for the visibility of the X-ray source for $\sim$60\% of the orbit.  This region then should be centered roughly on phase 0.05, measured along the equator from the impact point.  For the optical phase to lead the X-ray phase by 0.2, the optical region is centered on phase $\sim$0.25, which then implies the region has a full extent of $\sim$0.5 in phase, or half or more the circumference of the accreting white dwarf.

\item The fluid temperature must decrease from $\sim$50 eV ($\sim$500,000 K) to $\sim$25,000 K in the time it takes the equitorial flow to bring the newly-accreted fluid roughly half-way around the primary.  For a flow speed equal to the breakup velocity, this gives a time of 12 s, and an exponential cooling timescale of $\sim$4 s.  For a more reasonable flow speed that is one-tenth the breakup velocity, the cooling timescale is $\sim$40s.  Note that in this context ``cooling'' includes both radiative losses and entrainment of cooler gas from the surrounding photosphere.

\end{enumerate}

Of the three viable candidate models for the observed systems V407 Vul and HM Cnc, the DI model appears to require the least number of ``just so'' parameter choices, and to fit most easily the observational constraints.  The major issue with the DI model is the observed negative rate of period change, but as discussed above finite-temperature models of the secondary stars appropriate to these systems indicate overall contraction which can yield stable mass transfer and a decaying orbit for a sufficient span of time \citep{deloye06,dantona06}.  The polar model is then also not excluded by the observed $dP/dt$ rates.  Within the polar model the accretion impact spot is likely to be located within $\pi/2$ radians from the stellar line of centers, and this makes the observed phasing problematic.  The accretion stream trajectory in polars should be nearly vertical at impact, and because it is magnetically confined there should be a standing shock. The absence of a hard component in the spectrum as well as the absence of polarization are further long-standing problems with this model \citep{Cro2004}.  The unipolar inductor model appears to allow consistency with all of the observations -- including the reported detection of radio emission \citep{ram07} -- except the relative phasing of the optical and X-ray light curves.  In the UI model the $\sim$50 eV hot spots on the surface of the primary white dwarf are found along the meridian that passes through the line of centers, and the optical light curve results from the heated face of the secondary star.  This model nicely fit the old, incorrectly time-calibrated observations that the optical and X-ray light curves were anti-phased, but there appears to be no way for this model to explain the now currently accepted phase offset of 0.2 determined by \citet{Bar2007}. 

In future work, we plan to model the hydrodynamics of the impact region, using an implicit technique for estimating the radiation field and energy exchanges between SPH particles in the flux-limited diffusion approximation as described in \citet{whitehouse05}, and to use simple ray-trace concepts to estimate the emergent radiation as a function of inclination angle and orbital phase.

\acknowledgments

We thank Pasi Hakala for suggesting this project and Adela Kawka and Tom Marsh for useful discussions.  We also thank the referee for suggestions which improved the presentation of the results. This work was supported in part by grant AST-0205902 to The Florida Institute of Technology.

\clearpage

\begin{figure}
\plotone{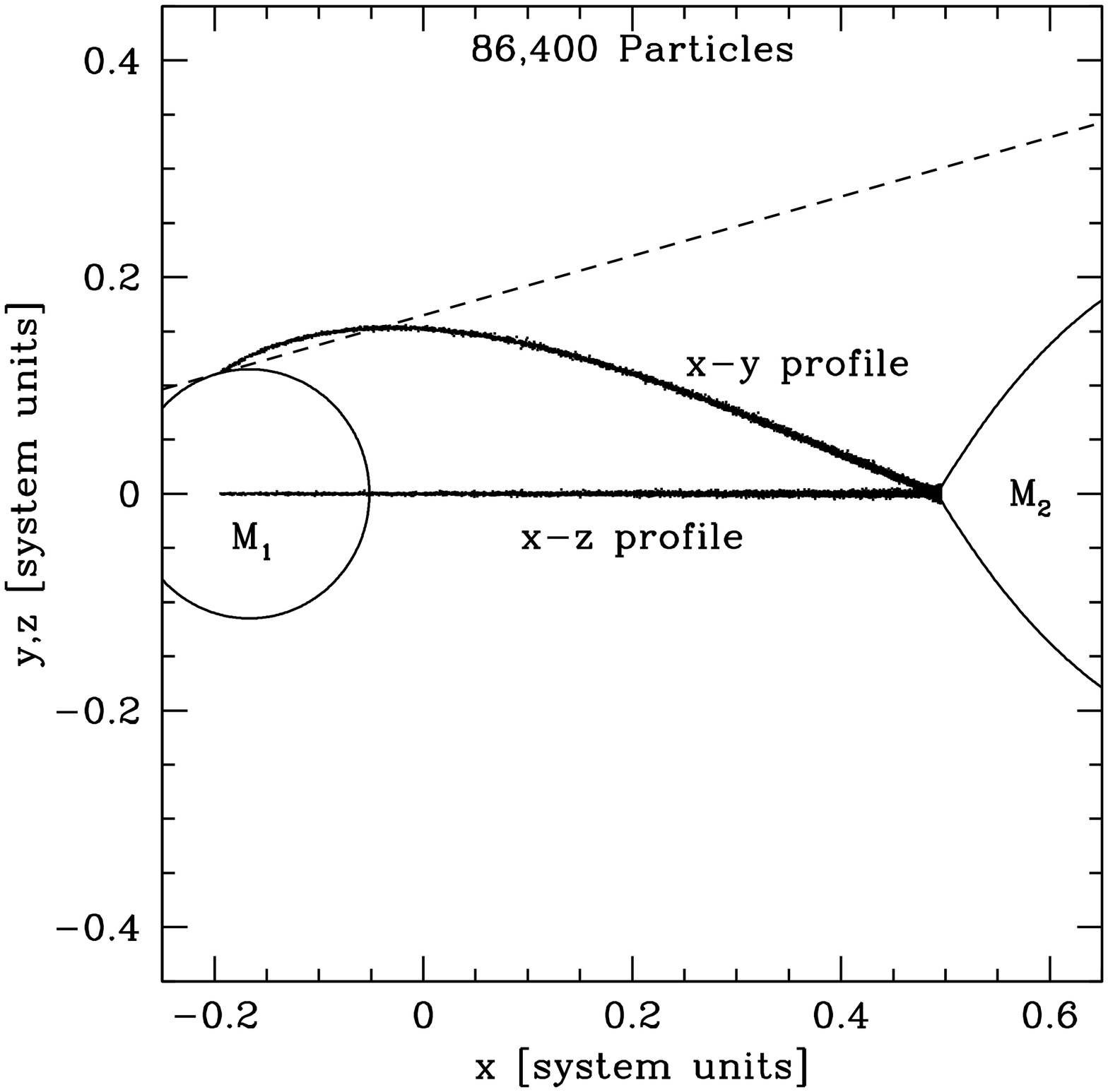}
\caption{Stream profiles in the planes parallel and perpendicular to the orbital (x-y) plane.  Also, the horizon of the impact spot is shown as the dotted line. Note that the altitude of the incoming accretion stream at the white dwarf surface is $\lesssim$20$^\circ$ above the local horizen.}
\end{figure}

\begin{figure}
\plotone{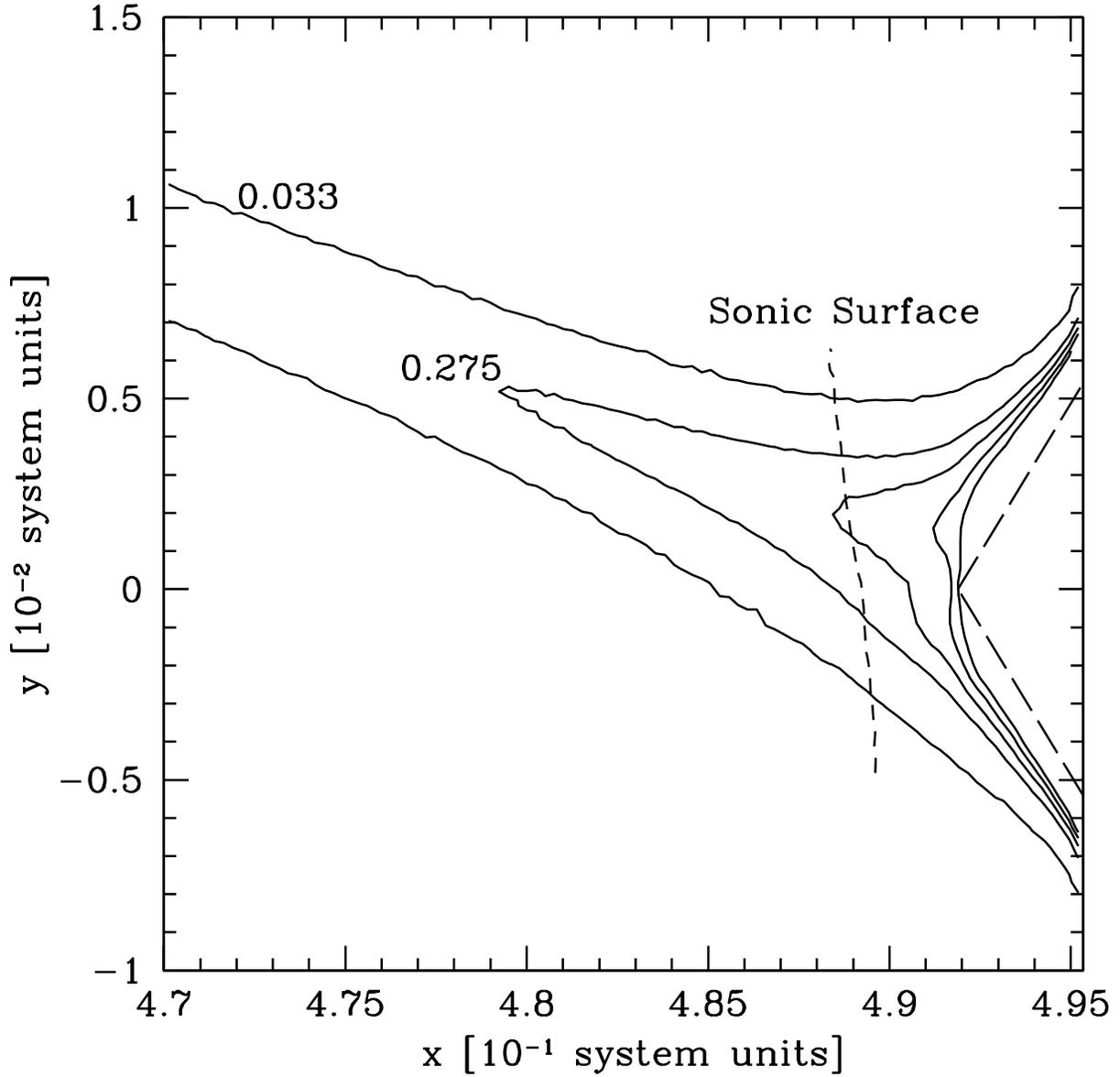}
\caption{Equidensity contours in the L1 region.  The density is normalized to unity on the contour that crosses over the L1 point.  Note the different scales on each axis.}
\end{figure}

\begin{figure}
\plotone{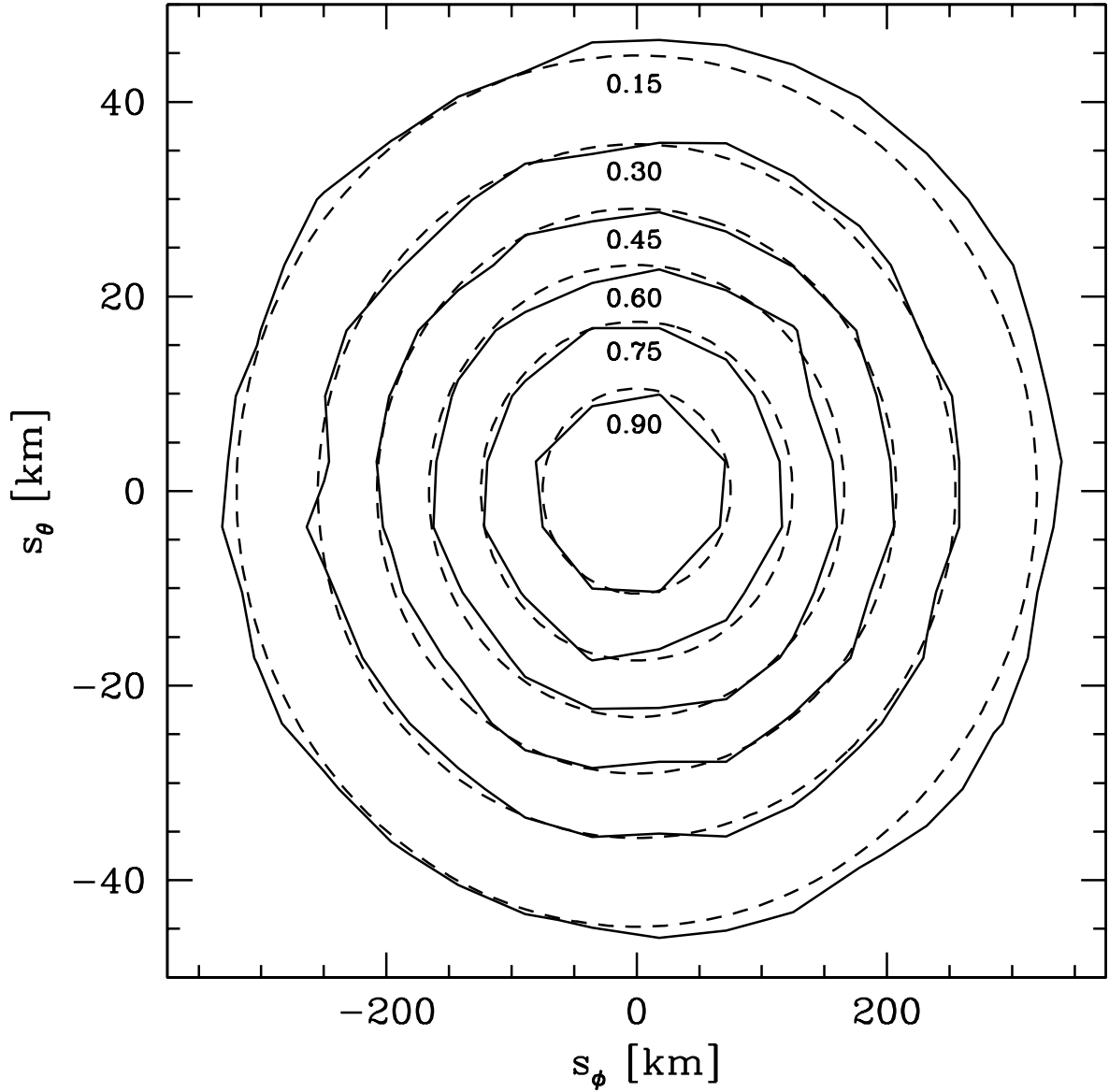}
\caption{Contours of constant mass flux at the impact spot.  The origin is set to the center of the fitted distribution (dashed lines).  The values shown have been normalized to unity at the central maximum value.  Note the different scales on each axis.}
\end{figure}

\begin{figure}
\plotone{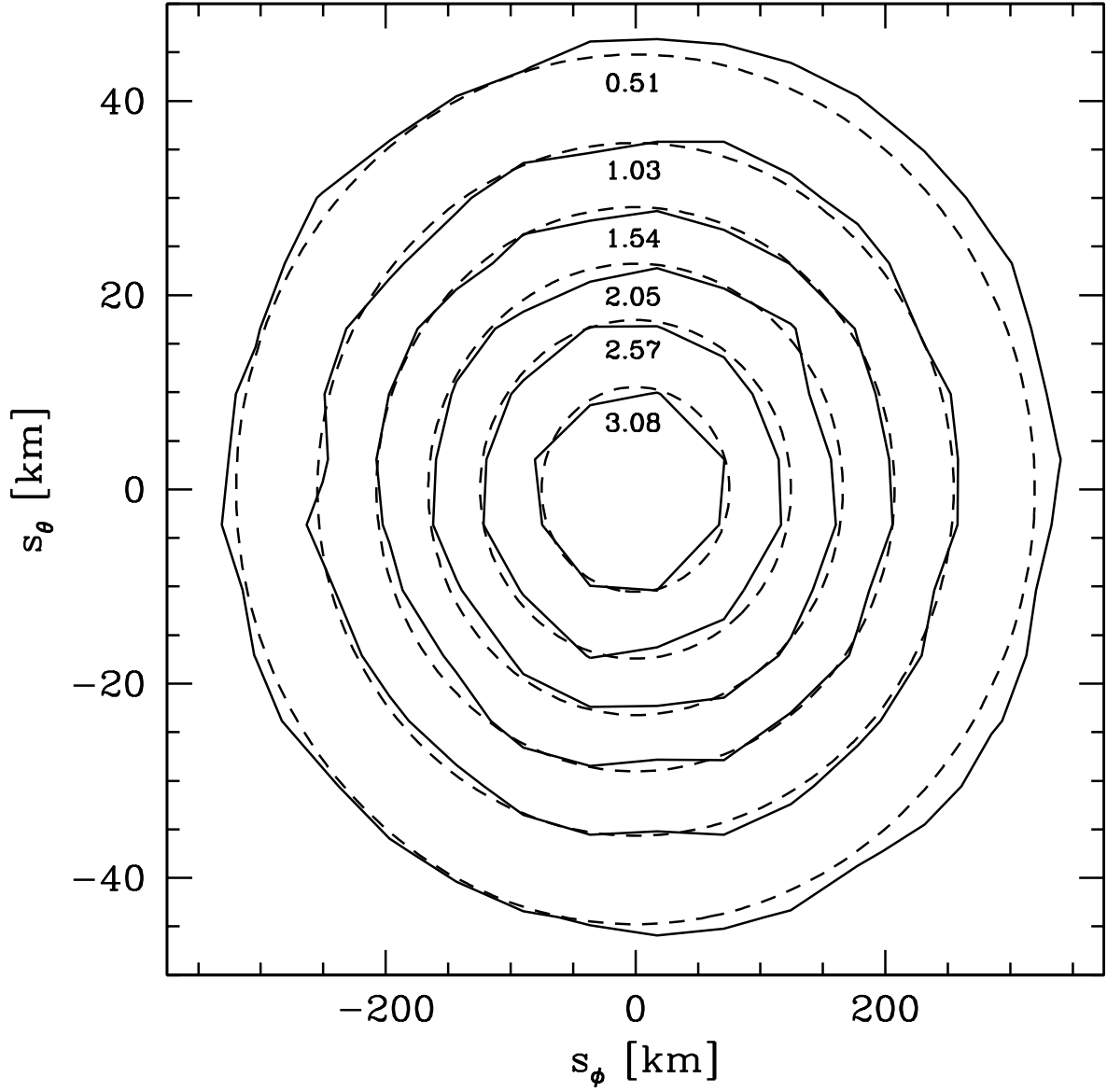}
\caption{Contours of constant ram pressure in the radial direction at the impact spot.  The values shown are in units of $10^{9}\rm\ dynes\rm\ cm^{-2}$.}
\end{figure}

\begin{figure}
\plotone{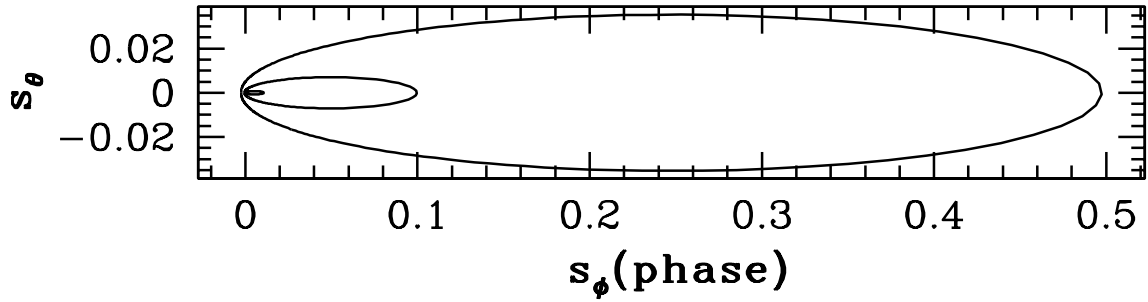}
\caption{Schematic diagram of a crude 3-zone model.  Assuming the impact spot, X-ray spot, and optical spot can all be characterized as uniformly-illuminated ellipses with the same eccentricity ($e=0.99$ is shown), and using the observational constraints, we find that the X-ray spot extends roughly 0.1 in phase, while the optical spot would extend $\sim$0.5 in phase, or half the circumference of the primary white dwarf.}
\end{figure}

\end{document}